\documentclass[letterpaper, 10 pt, conference]{ieeeconf}
\IEEEoverridecommandlockouts
\usepackage{graphicx} 
\usepackage{amsmath,amssymb,amsfonts,mathtools}
\usepackage{algorithm}
\usepackage{xcolor}
\graphicspath{ {images/} {plots/}}
\usepackage{enumerate}
\usepackage{cite}
\usepackage{accents}

\makeatletter
\let\NAT@parse\undefined
\makeatother
\usepackage{hyperref}
\hypersetup{
	colorlinks=true,
	linkcolor=blue,
	urlcolor=blue,
	citecolor=blue,
}

\newcommand{\optB}[0]{{\circ}}

\newcommand{\optD}[0]{{*}}

\newcommand{\xr}[0]{x}
\newcommand{\ur}[0]{u}
\newcommand{\xa}[0]{\xi}
\newcommand{\ua}[0]{\nu}
\newcommand{\zr}[0]{\boldsymbol{z}}
\newcommand{\za}[0]{\boldsymbol{\zeta}}
\newcommand{\OF}[0]{{O}}
\newcommand{\ellOF}[0]{{e}}
\newcommand{\SC}[0]{{S}}
\newcommand{\ellSC}[0]{{s}}
\newcommand{\app}[1]{\widetilde{#1}}
\newcommand{\rmFarg}[1]{\ensuremath{_{#1}(}}

\usepackage{balance}

\usepackage[noend]{algpseudocode}
\newtheorem{assumption}{Assumption}

\title{\LARGE \bf
Economic model predictive control for periodic operation:\\a quadratic programming approach
}
\author{Jose A.~Borja-Conde, Juan M.~Nadales,  Filiberto Fele, Daniel Limon
\thanks{This work was supported by grant PID2022-141159OB-I00 funded by MICIU/AEI/ 10.13039/501100011033 and by ERDF/EU. J.A.~Borja-Conde gratefully acknowledges support from grant FPU20/02023 funded by MU-Spain. F.~Fele gratefully acknowledges support from grant RYC2021-033960-I funded by MICIU/AEI/ 10.13039/501100011033 and European Union NextGenerationEU/PRTR.}
\thanks{J.A.~Borja-Conde, F.~Fele and D.~Limon are with the Department of Systems Engineering and Automation, University of Seville, Camino de los Descubrimientos s/n,  41092 Seville, Spain.  Juan M.~Nadales is with Earth-Life Science Institute, Tokyo Institute of Technology, Tokyo, Japan. (e-mail: {\tt \{jaborja,ffele,dlm\}@us.es}, {\tt nadales@elsi.jp}). }}

\begin{document}


\maketitle
\thispagestyle{empty}
\pagestyle{empty}

\begin{abstract}
Periodic dynamical systems, distinguished by their repetitive behavior over time, are prevalent across various engineering disciplines. In numerous applications, particularly within industrial contexts, the implementation of model predictive control (MPC) schemes tailored to optimize specific economic criteria was shown to offer substantial advantages. However, the real-time implementation of these schemes is often infeasible due to limited computational resources. To tackle this problem, we propose a resource-efficient economic model predictive control scheme for periodic systems, leveraging existing single-layer MPC techniques. Our method relies on a single quadratic optimization problem, which ensures high computational efficiency for real-time control in dynamic settings. We prove feasibility, stability and convergence to optimum of the proposed approach, and validate the effectiveness through numerical experiments.
\end{abstract}


\section{Introduction}

Periodic systems are characterized by repetitive behavior over time, meaning their dynamics, imposed constraints, or encountered disturbances recur at consistent intervals~\cite{bittanti2009periodic}. 
Examples of such systems can be found across a wide range of fields and applications.
Many nonlinear mechanical systems exhibit periodic orbits when subjected to external periodic forces~\cite{breunung2019does}. Periodic behavior is seen in various configurations of power electronic converters operating under different control strategies to manage power flow~\cite{van2002periodic}. Bioreactors utilized in the bioprocess industry demonstrate naturally periodic dynamics due to microbial growth rates and nutrient supply~\cite{amster2020existence}. The operation of heating, ventilation, and air conditioning (HVAC) systems is influenced by various typically periodic constraints, such as occupancy patterns, outdoor weather conditions, and time-of-day schedules~\cite{borja2024efficient}. In water distribution networks, periodic fluctuations in demand result from factors such as daily usage patterns and seasonal changes, requiring targeted management to ensure a consistent and efficient supply~\cite{berkel2018modeling}. In maritime transport systems, navigation is conditioned by the effect of the tide, which is of a periodic nature and conditions the temporary windows of access to river channels~\cite{nadales2024risk}.

Model predictive control (MPC) architectures have become a preferred choice for managing periodic systems.
This becomes evident from  the numerous research studies in the literature that explore this topic. To mention a few, \cite{gondhalekar2011mpc,limon2015mpc} proposed specialized MPC formulations for tracking periodic references, while \cite{gondhalekar2013least,pereira2016application} focused on the application aspects of these techniques in a wide variety of fields. 

The performance of industrial processes is typically evaluated using specific indices that reflect economic criteria. The objective of \emph{economic} MPC (E-MPC) formulations is to optimize system operation based on these economic indices~\cite{rawlings2017}. In this case, it has been shown that optimal performance is attained during steady-state operation when the system is strictly dissipative with respect to the economic cost function.~\cite{angeli2011average}. Moreover, under certain conditions---such as those present in periodic systems---economically optimal operation is achieved by following a periodic trajectory instead of maintaining a constant setpoint.
\cite{lee2001model}. 

To drive the system toward the economically optimal trajectory, a two-layer control framework is {typically} employed. {With} this approach, an optimal periodic reference is determined at the upper layer using a dynamic real-time optimizer (DRTO) that incorporates the system’s dynamics; at the lower layer, an MPC is used to track this optimal trajectory~\cite{wurth2011two}. {However, the differences between these two layers could lead to a loss of feasibility and stability. This has been addressed, for example, in~\cite{limon2012model}, where an approach to guarantee recursive feasibility and convergence despite variations in the economic cost function is proposed.}

Still, variations in the economic {criteria while the system is operating can} cause additional problems for the two-layer approach. {This is mainly because the optimal solution needs to} be recalculated in real time. However, the DRTO layer has a much longer computational time than the MPC, which could make online computation impossible, {leading to} a loss of optimality. In response to this issue, and following the idea proposed in~\cite{zanin2002integrating}, {a number of} studies have focused {on single-layer} E-MPC formulations, i.e., integrating the real-time DRTO and control stages into a unique layer. In the particular context of constrained \emph{periodic} linear systems, the authors of \cite{singleLayerPeriodicMPC} propose a single-layer architecture ensuring convergence to the optimal periodic trajectory. The formulation in \cite{muller2016economic} drops the requirement of the MPC terminal constraint under given dissipativity and controllability conditions, specifically focusing on scenarios where the optimal operating behavior is periodic.

A key difficulty with one-layer E-MPC approaches is their resolution in real time. {In} two-layer approaches, the lower layer is typically designed to be resource-efficient. {For the latter, it} is common to adopt a tracking MPC, {which} presents a quadratic program (QP) formulation {and} allows for the use of tailored solvers, enabling more {computationally} efficient implementations~\cite{IdrisEtAl2019CDC,krupa2020implementation}. {For the} specific context of periodic systems, the authors {of}~\cite{zanon2017periodic} focus on enhancing the performance of the tracking MPC layer to achieve local equivalence with the economic upper layer, although an upper layer to calculate the trajectories for tracking is still required. Note that this could be avoided if the E-MPC would directly present a QP {formulation; however, this is not common because of the inherently non-quadratic nature of economic objectives. An} approach to achieve the formulation of the E-MPC directly as a QP problem is presented in \cite{singleLayerPeriodicMPC}, although for non-periodic operation. This approach {is based} on a first-order Taylor approximation of the economic objective function, {guaranteeing} the convergence to the steady-state optimum (see also \cite{borja2024efficient}).

\subsection{Contribution}
The primary contribution of this paper is the development of a novel single-layer E-MPC algorithm specifically designed for periodic dynamical systems, which leverages the advantages of single-layer E-MPC methods and extends their application to periodic scenarios. Our proposal builds upon the gradient-based approximation in \cite{singleLayerPeriodicMPC} where, at each sampling period, the solution of the E-MPC control problem is attained by solving a single QP problem. This provides our algorithm with high computational efficiency and positions it as a viable option for expanding the scope of embedded MPC applications to include periodic economic MPC control problems.

The rest of the paper is structured as follows. Preliminary sections~\ref{sec:problemformulation} and~\ref{sec:economicMPCforChangingPeriodicOperation} present the control objective and summarize the periodic E-MPC formulation in \cite{singleLayerPeriodicMPC}. The main contribution of the paper is found in {Sections~\ref{sec:gradientbasedapproach} and \ref{sec:convergence},} where our QP-based single-layer periodic E-MPC algorithm is illustrated. Section~\ref{sec:casestudy} presents a numerical example that demonstrates the performance of our approach.

\section{Problem Formulation}\label{sec:problemformulation}
Consider a linear time-varying system whose dynamics are described by the following discrete-time equation:
\begin{equation} \label{eq:systemmodel}
{\xr}_{{k+}1} = {f}\rmFarg{{k}}{\xr}_{{k}}, {\ur}_{{k}}
) = {A}_{k} {\xr}_{k} + {B}_{k}{\ur}_{k},
\end{equation}
\noindent where ${\xr}_{k}\in \mathbb{R}^{n}$ and ${\ur}_{k}\in \mathbb{R}^{m}$ are the state of the system and the control input, respectively, at time instant $k$.
It is assumed that the pair $({A}_{k},{B}_{k})$ is controllable and that the evolution of ${A}_{k}$ and ${B}_{k}$
for $k\in \mathbb{N}$ is perfectly known. 

We consider possibly joint constraints on the state of the system and control input as $({\xr}_{{k}}, {\ur}_{{k}}) \in \mathcal{Z}_{k} \subseteq \mathbb{R}^{n+ m}$, where $\mathcal{Z}_{k}\subseteq \mathbb{R}^{n+ m}$ is a time-varying closed convex polyhedron that contains the origin in its interior.

The stage cost  $\mathcal{\ell}^\ellOF\rmFarg{{k}}{\xr}_{{k}},{\ur}_{{k}},{p})$ evaluates the economic performance associated with the system  at time instant $k$. This, in addition to the state and control input, depends on an exogenous parameter $p$, which is allowed to be time-varying (it could also indicate the switching between different economic criteria); no assumption is made on the policy governing the behavior of this parameter. {For the sake of a compact notation, we refer to ${p}_{k}$ as ${p}$ in the rest of the paper.} 

\begin{assumption}\label{ass:periodic}
    The system is periodic, with a periodicity of $T \in \mathbb{Z}_{>0}$ sampling periods, i.e, at any time instant $k$ it holds that
    \begin{subequations}
    \begin{align}
     &{A}_{k}  = {A}_{k+T},\ {B}_{k} = {B}_{k+T},\ \mathcal{Z}_{k} = \mathcal{Z}_{k+T},\\
     &\mathcal{\ell}^\ellOF\rmFarg{{k}}{\xr},{\ur},{p})  = \mathcal{\ell}^\ellOF\rmFarg{k+T}{\xr},{\ur},{p}), \ \forall ({\xr}, {\ur}) \in \mathcal{Z}_{k},\forall p.
    \end{align}
    \end{subequations}
\end{assumption}

The performance of the system is assessed by calculating the average of the economic cost derived from the closed-loop system trajectories, i.e.,
\begin{equation}
\label{eq:L_infinito}
    \underset{M\to\infty}{\lim} \frac{1}{{M}}\sum_{{j=}0}^{{M}-1}\mathcal{\ell}^\ellOF\rmFarg{{j}}{\xr}_{{j}},{\ur}_{{j}},{p}).
\end{equation}

From this point forward, to keep notation compact, we may omit the subscript $k$ when denoting quantities relative to instant $k$, if it is understood from the context.

Since the system has a periodic behavior as per Assumption~\ref{ass:periodic}, from \cite[Theorem 1]{singleLayerPeriodicMPC} we have that the trajectory which minimizes the cost in~\eqref{eq:L_infinito} is also periodic. This optimal trajectory, which we denote $({\xr}^\optB,\mathbf{\ur}^\optB)$, can be obtained by solving the following DRTO problem:
\begin{subequations}\label{eq:D}
\begin{align} 
    ({\xr}^\optB,\mathbf{\ur}^\optB) = \arg \, \mathbb{D}&\rmFarg{{k}}{p}) :\ \underset{{\xr}_{0},\mathbf{\ur}}{\min} \  \sum_{{j=}0}^{T-1}\mathcal{\ell}^\ellOF\rmFarg{{k+j}}{\xr}_{{j}},{\ur}_{{j}},{p}) \label{eq:Da}\\
        \text{s.t.}\quad& {\xr}_{{j+}1}={f}\rmFarg{{k+j}}{\xr}_{{j}},{\ur}_{{j}}),\; {{\xr}_{T}={\xr}_{0},} \\
        &({\xr}_{{j}},{\ur}_{{j}}) \in \mathcal{Z}_{{k+j}}\,,\  \ \forall j \in \mathbb{Z}_{[0,T-1]},
\end{align}
\end{subequations}
where $\mathbf{\ur} \coloneqq [{\ur}_{0}, \dots ,{\ur}_{T-1}]\in\mathbb{R}^{mT}$, {and $\mathbb{Z}_{[0,T-1]}$ is the set of integers from $0$ to $T-1$.}
\begin{assumption}\label{ass:conv_stage_cost}
The function $\mathcal{\ell}^\ellOF_{{k}}$ is convex in $({\xr}_{{k}}, {\ur}_{{k}})$ and lower-bounded for all value of $({\xr}_{{k}}, {\ur}_{{k}}) \in \mathcal{Z}_{{k}} $, $p$ and ${k}$.
In addition, there {exists $\alpha \in \mathcal{K}_\infty$} such that
\begin{multline*}
\ \mathcal{\ell}^\ellOF\rmFarg{{k+j}}{\xr}_{{j}},{\ur}_{{j}},{p}) - \mathcal{\ell}^\ellOF\rmFarg{{k+j}}{\xr}^\optB_{{j}},{\ur}^\optB_{{j}},{p}) \geq  \\
\alpha(\lVert ({\xr}_{{j}},{\ur}_{{j}}) - ({\xr}^\optB_{{j}},{\ur}^\optB_{{j}})  \rVert),  \ \forall {k},{j}.
\end{multline*}
\end{assumption}

We note that $({\xr}^\optB,\mathbf{\ur}^\optB)$ is unique under Assumption~\ref{ass:conv_stage_cost}.
 
\begin{assumption}
    \label{ass:lipschitz}
    The gradient of {$\mathcal{\ell}^\ellOF\rmFarg{{k}}{\xr},{\ur},{p})$ with respect to $(\xr,\ur)$ is $\rho$-Lipschitz continuous, i.e.,
    \begin{multline*}
    	\lVert { \nabla_{(x,u)}\mathcal{\ell}^\ellOF\rmFarg{{k}} x,u,{p}) } - { \nabla_{(v,w)}\mathcal{\ell}^\ellOF\rmFarg{{k}} v, w,{p}) } \rVert \leq \\
    	\rho \lVert (x,u) - (v,w)\rVert, \ \forall p,
    \end{multline*}
for all $(x,u),(v,w)\in \mathcal{Z}_{{k}}$ and some $\rho>0$.}
\end{assumption}

In the next section, the goal is to design a controller that---for any given value of $p$---stabilizes the system and steers it to the  optimal trajectory $({\xr}^\optB(p),\mathbf{\ur}^\optB(p))$, while also satisfying the constraints during transients.

\section{Economic MPC for Changing\\Periodic Operation}
\label{sec:economicMPCforChangingPeriodicOperation}
{First,} we summarize the MPC control formulation proposed in \cite{singleLayerPeriodicMPC} that addresses the aforementioned control objective. To keep notation {compact, we write  ${\zr}\coloneqq (\xr,\mathbf{\ur})\in\mathbb{R}^{n+mN}$ when designating the pair composed of a given initial state ${\xr}$ and a sequence of $N$ control inputs $\mathbf{\ur}$.} 

One of the main building blocks of our E-MPC control scheme is a MPC controller which tracks periodic references~\cite{singleLayerPeriodicMPC}.
To this aim, we introduce an artificial periodic trajectory $\za=({\xa},\boldsymbol{\ua})\in \mathbb{R}^{n+m T}$ as a decision variable: {the trajectory} is determined by an initial state $\xa$ and a sequence $\boldsymbol{\ua}\coloneqq  [{\ua}_{0}, \dots ,{\ua}_{T-1}]\in\mathbb{R}^{mT}$ of $T\geq N$ control inputs $\ua$.
{The deviation} between the predicted and artificial trajectories is penalized through the stage cost $\SC:\mathbb{R}^{n+m N}\times \mathbb{R}^{n+m T}\rightarrow \mathbb{R}$, defined as
\begin{equation}\label{eq:S}
\SC({\zr},{\za}) = \sum_{{i=}0}^{N-1}
\mathcal{\ell}^\ellSC({\xr}_{{i}} - {\xa}_{{i}},{\ur}_{{i}} - {\ua}_{{i}}) ,
\end{equation}
where $\mathcal{\ell}^\ellSC:\mathbb{R}^{n}\times\mathbb{R}^{m}\rightarrow \mathbb{R}$ is defined as
\begin{displaymath}
    \mathcal{\ell}^\ellSC(v,w) = \lVert v  \rVert^2_{Q}+\lVert w  \rVert^2_{R},
\end{displaymath}
{and $Q,R$ are positive definite matrices.}
\noindent The economic cost of the artificial trajectory ${\za}$ along the horizon $T$ is evaluated by the offset cost function  $\OF_{{k}}\colon\mathbb{R}^{n+m T}\rightarrow \mathbb{R}$, defined as:
\begin{equation}\label{eq:O}
\OF\rmFarg{{k}}\boldsymbol{\za},{p}) = \sum_{{j=}0}^{T-1}\mathcal{\ell}^\ellOF\rmFarg{{k+j}}{\xa}_{{j}},{\ua}_{{j}},{p}),
\end{equation}
where $\ell^e_{k+j}$ is as in \eqref{eq:Da}.
Combining \eqref{eq:S} and \eqref{eq:O}, the cost function of the proposed single-layer MPC problem, ${V}_{k}:\mathbb{R}^{n+m N}\times \mathbb{R}^{n+m T}\rightarrow \mathbb{R}$, is defined as:
\begin{equation}\label{eq:orig_cost}
    {V}\rmFarg{{k}}{\zr},{\za},{p}) \coloneqq \SC({\zr},{\za}) + \OF\rmFarg{{k}}{\za},{p}).
\end{equation}

Then, for a given state ${\xr}$ and parameter ${p}$ at time $k$, the single-layer MPC control problem is formulated as:
\begin{subequations}
\label{eq:P}
\begin{align}
\mathbb{P}\rmFarg{{k}}{\xr},{p}) :\ & \underset{{\zr},{\za}}{\min} \  {V}\rmFarg{{k}}{\zr},{\za},{p}) \label{}\\
\label{eq:P_2_ini}
    \text{s.t.} \quad 
    &{\xr}_{{i+}1}={f}\rmFarg{{k+i}}{\xr}_{{i}},{\ur}_{{i}}), \\
    &{\xa}_{{j+}1}={f}\rmFarg{{k+j}}{\xa}_{{j}},{\ua}_{{j}}), \\
    &({\xr}_{{i}},{\ur}_{{i}}) \in \mathcal{Z}_{{k+i}},\, \ \forall i \in \mathbb{Z}_{[0,N-1]},\\
    &({\xa}_{{j}},{\ua}_{{j}}) \in \mathcal{Z}_{{k+j}},\, \ \forall j \in \mathbb{Z}_{[0,T-1]},\\
    &{\xr}_{0}={\xr}, \ { {\xa}_{T}={\xa}_{0}, } \ {\xr}_{N}={\xa}_{N}. \label{eq:P_2_end}
\end{align}
\end{subequations}

Problem \eqref{eq:P} is referred to as single-layer economic MPC as it integrates into a unique optimization problem the two layers of the canonical E-MPC formulation, i.e., the upper-layer DRTO and the lower-layer tracking MPC.

    From \cite[Theorem~3]{singleLayerPeriodicMPC} we have that---for any feasible initial state---system \eqref{eq:systemmodel} driven by the optimal feedback policy derived as the solution of \eqref{eq:P} is recursively feasible, and the optimal trajectory $({\xr}^\optB,\mathbf{\ur}^\optB)$ is an asymptotically stable equilibrium for the closed-loop system.



\section{Gradient-Based Approach}
\label{sec:gradientbasedapproach}
We note that problem~\eqref{eq:P} is convex but is not, in general, a
QP problem---which can make its implementation difficult.
In this paper, inspired by the ideas in~\cite{gradientBasedSingleLayerMPC}, we present a gradient-based approximated formulation of the E-MPC for changing periodic operation; this approach relies upon a first-order approximation of the economic objective function, allowing to reduce the control problem  complexity to that of a QP program, while maintaining the same properties as the original control {problem}.    

Let the approximated single-layer MPC cost function,  $\app{V}_{k}:\mathbb{R}^{n+m N}\times \mathbb{R}^{n+m T}\times \mathbb{R}^{n+m T}\rightarrow \mathbb{R}$,  be defined as:
\begin{equation}\label{eq:approx_cost}
\app{V}\rmFarg{{k}}{\zr},{\za},\hat{{\za}},{p})= \SC({\zr},{\za})+\app{\OF}\rmFarg{{k}}{\za},\hat{{\za}},{p}),
\end{equation}
where $\hat{\za}$ is a given feasible trajectory about which we take the first-order Taylor approximation of the economic cost, and $\app{\OF}_{k}:\mathbb{R}^{n+m T}\times \mathbb{R}^{n+m T}\rightarrow \mathbb{R}$ is the approximated offset cost, given by the following quadratic cost function:
\begin{multline}
\label{eq:O_app}
        \app{\OF}\rmFarg{{k}}{\za},\hat{{\za}},{p}) =\sum_{{j=}0}^{T-1}\Bigr( \mathcal{\ell}^\ellOF\rmFarg{{k+j}}\hat{{\xa}}_{{j}},\hat{{\ua}}_{{j}},{p})\\
        +{ \nabla_{(\xa,\ua)}\mathcal{\ell}^\ellOF\rmFarg{{k+j}}\hat{{\xa}}_{{j}},\hat{{\ua}}_{{j}},{p}) }  \cdot\left[{({\xa}_{{j}},{\ua}_{{j}})-(\hat{{\xa}}_{{j}},\hat{{\ua}}_{{j}})}\right]\\ +\frac{\rho}{2}\left\lVert{({\xa}_{{j}},{\ua}_{{j}})-(\hat{{\xa}}_{{j}},\hat{{\ua}}_{{j}})}\right\rVert^2\, \Bigr), 
\end{multline}
\noindent where $\rho$ is the Lipschitz constant for $\ell^e_k$ as per Assumption~\ref{ass:lipschitz}, so it holds that
\begin{multline}
\label{eq:upperbound}
\mathcal{\ell}^\ellOF\rmFarg{{k}}{{\xa}},{{\ua}},{p}) \leq \mathcal{\ell}^\ellOF\rmFarg{{k}}\hat{{\xa}},\hat{{\ua}},{p})
+{ \nabla_{(\xa,\ua)}\mathcal{\ell}^\ellOF\rmFarg{{k}}\hat{{\xa}},\hat{{\ua}},{p}) }  \cdot\left[{({\xa},{\ua}){-}(\hat{{\xa}},\hat{{\ua}})}\right]\\ +\frac{\rho}{2}\left\lVert{({\xa},{\ua})-(\hat{{\xa}},\hat{{\ua}})}\right\rVert^2, \ \forall k.
\end{multline}
\noindent Note that $\app{\OF}$ is an upper-bound of $\OF$. Then, replacing $\OF$ by $\app{\OF}$ in $\mathbb{P}$, the approximated MPC problem is obtained. The optimal solution for this is:
\begin{equation}\label{eq:control_input}
({\zr}^\optD,{\za}^\optD) =  \arg\app{\mathbb{P}}\rmFarg{{k}}{\xr},\hat{{\za}},{p}),
\end{equation}
with
\begin{subequations}
\label{eq:P_app}
\begin{align}
\app{\mathbb{P}}\rmFarg{{k}}{\xr},\hat{{\za}},{p}) & : \quad {\underset{{\zr},{\za}}{\min} \  \app{V}\rmFarg{{k}}{\zr},{\za},\hat{{\za}},{p}) }\\ 
&\text{s.t.} \quad  \text{\eqref{eq:P_2_ini} to \eqref{eq:P_2_end}}.
\end{align}
\end{subequations}

Then, the MPC feedback law is obtained as
\begin{equation*}
    \app{\kappa}\rmFarg{{k}}x,\hat{\za},p) =u^\optD_0,
\end{equation*}
i.e., it is given by the application of the first element of the control sequence $\mathbf{\ur}^\optD$---part of the solution ${\zr}^\optD$ of \eqref{eq:P_app} at each time $k$---in a receding-horizon fashion.

\subsection{Proposed algorithm}
Next, we introduce our single-layer E-MPC algorithm inspired by Newton's optimization method, where the linearization point is selected to be the optimal artificial trajectory of the optimization problem solved at the previous time step. The resulting procedure is described in Algorithm~\ref{alg:approximatedMPC}. It can be noticed from \eqref{eq:control_input} that $u^\optD_0$ depends on the exogenous parameter $\hat{{\za}}$. A suitable initial value for $\hat{{\za}}$ could be obtained by {solving~\eqref{eq:P_app} first with} $\app{\OF}$ as any quadratic offset cost function (since this initial point does not need to be optimal but only feasible).

\begin{algorithm}[t]
\caption{Approximated single-layer periodic E-MPC}
\label{alg:approximatedMPC}
\textbf{Require:} Initial feasible $\hat{{\za}}=(\hat{{\xa}}, \hat{\boldsymbol{\ua}})$. 
\\At each sampling time ${k}\geq 0$:
\begin{algorithmic}[1]
  \State \textbf{Read}: ${\xr}$, $p$
  \State \textbf{Solve QP}: $({\zr}^\optD,{\za}^\optD)=\arg\app{\mathbb{P}}\rmFarg{{k}}{\xr},\hat{{\za}},{p})$
  \State \textbf{Apply}: ${\ur}^\optD_{0}$
  \State \label{algStep:approximatedMPCstep4}\textbf{Update}: $\hat{\za} \colon \ \hat{{\xa}} \leftarrow {\xa}_1^\optD ,\, \hat{\boldsymbol{\ua}} \leftarrow \begin{bmatrix}{\ua}^\optD_{1},\dots,{\ua}^\optD_{T\text{-}1},{\ua}^\optD_{0}\end{bmatrix}$
\end{algorithmic}
\end{algorithm}

\section{Stability and Convergence Analysis}
\label{sec:convergence}
Algorithm~\ref{alg:approximatedMPC} implements \eqref{eq:P_app} {in a receding horizon fashion}. In the remainder of this section, we show that its solution achieves  the same desirable closed-loop properties of the original (non-QP) E-MPC $\mathbb{P}$, namely, (i) for any feasible initial state the optimization problem is recursively feasible, and (ii) the optimal trajectory of the system is asymptotically stable for the controlled system. These properties are formally presented next.

\begin{assumption}[Controllability]
    \label{ass:controllability_in_c_steps}
    For integers $0 \leq i < j$ let
    \begin{displaymath}
    \Psi(j,i) = \left ( A_{j}\cdot A_{j-1} \cdots A_{i+1} \right ) \cdot B_i. 
    \end{displaymath}
    \noindent 
    Then, let the controllability matrix $\Gamma$ be defined as
    \begin{multline*}
        \Gamma_k(c)=[\Psi(k+c,k), \Psi(k+c,k+1), \dots\\
        \dots,\Psi(k+c,k+c-1),B_{k+c}].
    \end{multline*}
    \noindent We assume that
    \begin{equation*}
    \exists c \in [0,T-1]\colon \mathrm{rank}\left(\Gamma_k(c)\right) = n, \, \forall k.
    \end{equation*}
    
\end{assumption}

\subsection{Recursive feasibility:}
    In~\cite{singleLayerPeriodicMPC}, it is proved that $\mathbb{P}$ is recursively feasible. Since the constraints in $\mathbb{P}$ are the same as those in $\app{\mathbb{P}}$, this is recursively feasible irrespective of the selected values of $\hat{\za}$.

\subsection{Stability and convergence:}
    From instant $k$, assume a feasible periodic trajectory $\hat{{\za}}_{k}$. From the feasible state ${\xr}_{k}$ and considering $\hat{{\za}}_{k}$ for the linearization of $\app{\mathbb{P}}$, the solution to $\app{\mathbb{P}}\rmFarg{{k}}{\xr_{k}},\hat{{\za}_{k}},{p})$ would be $({\zr}^\optD_{k},{\za}^\optD_{k})$. Thus, the cost associated to the system {at} instant $k$ would be $\app{V}\rmFarg{{k}}{\zr}_{k}^\optD,{\za}_{k}^\optD,\hat{{\za}}_{k},{p})$. First it is proved that this is a decreasing function.

    For instant $k+1$, {we propose a solution} for this new instant, denoted as {${\zr}_{k}^{{+}}\coloneqq ({\xr}^{{+}}_{k,0},\mathbf{\ur}_{k}^{{+}})$,} such that\footnote{{Here, we denote by $\xr_{k,i}$ the value of $\xr$ at time $k+i$ predicted from the initial condition ${\xr_k}$.}}
    \begin{equation}
    \nonumber
    {\xr}^{{+}}_{k,0} = {\xr}^\optD_{k,1} \,, \quad\mathbf{\ur}_{k}^{{+}} = \begin{bmatrix}{\ur}^\optD_{k,1},\dots,{\ur}^\optD_{k,N{-}1},{\ua}^\optD_{k,N}\end{bmatrix}.
    \end{equation}
    \noindent This solution is feasible, as proven in~\cite[Theorem~3]{singleLayerPeriodicMPC}. Next, according to step~\ref{algStep:approximatedMPCstep4} in Algorithm~\ref{alg:approximatedMPC}, the feasible trajectory for the linearization in this instant $k+1$, denoted as $\hat{{\za}}_{k+1}$, is obtained by shifting ${\za}_{k}^\optD$ as:
    \begin{equation}
    \label{eq:hat_za_k+1}
    \hat{{\xa}}_{k+1,0} = {\xa}^\optD_{k,1} \,, \quad\hat{\boldsymbol{\ua}}_{k+1} = \begin{bmatrix}{\ua}^\optD_{k,1},\dots,{\ua}^\optD_{k,T{-}1},{\ua}^\optD_{k,0}\end{bmatrix}.
    \end{equation}
    \noindent Denote
    \begin{displaymath}
    \Delta\app{V}
    = {\app{V}\rmFarg{{k+}1}{\zr}_{k}^{{+}},\hat{{\za}}_{k+1},\hat{{\za}}_{k+1},{p}) }
    -
    \app{V}\rmFarg{{k}}{\zr}_{k}^\optD,{\za}_{k}^\optD,\hat{{\za}}_{k},{p}).
    \end{displaymath}
    \noindent {For fixed $p$, we have that:}
    \begin{multline}
    \label{eq:deltaV_approximated}
    \Delta\app{V}
    = 
    \underbrace{\app{V}\rmFarg{{k+}1}{\zr}_{k}^{{+}},\hat{{\za}}_{k+1},\hat{{\za}}_{k+1}) 
    -
    \app{V}\rmFarg{{k+}1}{\zr}_{k}^{{+}},\hat{{\za}}_{k+1},\hat{{\za}}^{{+}}_{k})}_{{\Delta{V}_1} }  \\    
    +
    \underbrace{\app{V}\rmFarg{{k+}1}{\zr}_{k}^{{+}},\hat{{\za}}_{k+1},\hat{{\za}}^{{+}}_{k})  
    -
    \app{V}\rmFarg{{k}}{\zr}_{k}^\optD,{\za}_{k}^\optD,\hat{{\za}}_{k})}_{{\Delta{V}_2}},
    \end{multline}
    \noindent where $
        \label{eq:hat_za_k_+}
        \hat{{\xa}}_{k,0}^{{+}} = \hat{{\xa}}_{k,1} \,, \quad\hat{\boldsymbol{\ua}}_{k}^{{+}}
        =
        \begin{bmatrix}\hat{{\ua}}_{k,1},\dots,\hat{{\ua}}_{k,T{-}1},\hat{{\ua}}_{k,0}\end{bmatrix}.
    $
    
    The first term of the right hand side of the equation can be expanded as: 
    \begin{multline}
        \Delta V_1 = 
        \sum_{{j=}0}^{T-1}\mathcal{\ell}^\ellOF\rmFarg{{k+j}}{\xa}^{\optD}_{{k,j}},{\ua}^{\optD}_{{k,j}},{p})
        -
        \Bigr( \sum_{{j=}0}^{T-1}\Bigr( \mathcal{\ell}^\ellOF\rmFarg{{k+j}}\hat{{\xa}}_{{k,j}},\hat{{\ua}}_{{k,j}},{p})\\
        +
        { \nabla_{(\xa,\ua)}\mathcal{\ell}^\ellOF\rmFarg{{k+j}}\hat{{\xa}}_{{k,j}},\hat{{\ua}}_{{k,j}},{p}) }\cdot\left[{({\xa}^{\optD}_{{k,j}},{\ua}^{\optD}_{{k,j}})-(\hat{{\xa}}_{{k,j}},\hat{{\ua}}_{{k,j}})}\right]\\
        +
        \frac{\rho}{2}\left\lVert{({\xa}^{\optD}_{{k,j}},{\ua}^{\optD}_{{k,j}})-(\hat{{\xa}}_{{k,j}},\hat{{\ua}}_{{k,j}})}\right\rVert^2\, \Bigr).
    \end{multline}
    \noindent From \eqref{eq:upperbound}, it holds for all $j$ that
    \begin{multline}\label{eq:lemma1a}
                {\mathcal{\ell}^\ellOF\rmFarg{{k+j}}{\xa}_{{j}},{\ua}_{{j}},{p}) - \Bigr(                \mathcal{\ell}^\ellOF\rmFarg{{k+j}}\hat{{\xa}}_{{j}},\hat{{\ua}}_{{j}},{p})+ }\\
                { \nabla_{(\xa,\ua)}\mathcal{\ell}^\ellOF\rmFarg{{k+j}}\hat{{\xa}}_{{j}},\hat{{\ua}}_{{j}},{p}) } \cdot\left[{({\xa}_{{j}},{\ua}_{{j}})-(\hat{{\xa}}_{{j}},\hat{{\ua}}_{{j}})}\right]\\
                +\frac{\rho}{2}\left\lVert{({\xa}_{{j}},{\ua}_{{j}})-(\hat{{\xa}}_{{j}},\hat{{\ua}}_{{j}})}\right\rVert^2\Bigr) \leq 0.
    \end{multline}
    {The inequality still} holds by taking the sum from ${j}=0$ to $T-1$ of both left- and right-hand side terms in \eqref{eq:lemma1a}. Therefore, $\Delta V_1 \leq 0$. In addition, it is proven in~\cite[Theorem~3]{singleLayerPeriodicMPC} that the second term of the left hand side of equation~\eqref{eq:deltaV_approximated} is such that
    \begin{equation}
        \Delta V_2 \leq -\mathcal{\ell}^\ellSC( {\xr}^\optD_{k,0} - {\xa}^\optD_{k,0},{\ur}^\optD_{k,0} - {\ua}^\optD_{k,0}).
    \end{equation}
    \noindent {From the above, and recalling} ${\xr}^\optD_{k,0} = {\xr}_k$, we have that
    \begin{equation}
        \Delta\app{V}^\optD \leq 
        \Delta\app{V} \leq -\mathcal{\ell}^\ellSC( {\xr}_{k} - {\xa}^\optD_{k,0},{\ur}^\optD_{k,0} - {\ua}^\optD_{k,0}),
    \end{equation}
    \noindent {with $\Delta\app{V}^\optD = \app{V}\rmFarg{{k+}1}{\zr}_{k+1}^{\optD},\za_{k+1}^\optD,\hat{{\za}}_{k+1},{p}) - \app{V}\rmFarg{{k}}{\zr}_{k}^\optD,{\za}_{k}^\optD,\hat{{\za}}_{k},{p})$.} From this, it is inferred that the system converges to an admissible periodic trajectory ${\zr}_{\infty}={\za}_{\infty}$ and that at ${\zr}_{\infty}$ the optimal solution is such that ${\zr}^\optD_{\infty}= {\zr}_{\infty}$, ${\za}^\optD_{\infty}= {\za}_{\infty}$.

    The optimal cost function at $({\zr}_{\infty},{\za}_{\infty})$ is $\app{V}\rmFarg{{k}}{\zr}_{\infty},{\za}_{\infty},{\za}_{\infty},{p})=\OF\rmFarg{{k}}{\za}_{\infty},{p})$. Considering that $\OF$ is convex, based on~\cite{singleLayerPeriodicMPC} it can be proved that the system converges to a feasible periodic trajectory such that
\begin{equation}
    {\zr}^\optD_{\infty} \rightarrow {\za}^\optD_{\infty} \rightarrow \arg  \underset{{\za}}{\min} \  \OF\rmFarg{{k}}{\za},{p}),
\end{equation}
\noindent i.e., the solution to $\mathbb{D}$. Thus, ${\zr}^\optD_{\infty} \rightarrow {\za}^\optB$.

\section{Numerical example}\label{sec:casestudy}
The proposed approach is implemented in a linearized ball and plate system, where the purpose is to control a ball rolling on a plate. Note that, in this numerical example, the periodicity of the system is not directly induced by the system behaviour, but by a periodic reference trajectory to be followed by the ball. The position of the ball is denoted as $[y_1,y_2]$. The controlled angles of the plate are $[\theta_1,\theta_2]$. A discrete-time linear system in the form of \eqref{eq:systemmodel} is derived by setting the equilibrium point as the origin for all states and inputs, with a sampling time of $0.05$ s, as in \cite{moreno2010,wang2014}. The input of the system is the angular acceleration $u= [\ddot{\theta}_1, \ddot{\theta}_2]^\top $.

The state vector $x$ is defined by the position and velocity of both the ball and angles, i.e., $x = [y_1, \dot{y}_1, \theta_1, \dot{\theta}_1, y_2, \dot{y}_2, \theta_2, \dot{\theta}_2]^\top$.
The system constraints are:
\begin{subequations}
\begin{align}
    |y_1|+|y_2| \leq 6 \text{ cm}, \\
    |\theta_i| \leq \frac{\pi}{2} \text{ rad}, \quad |\ddot{\theta}_i| \leq 110 \text{ rad/s}^2,  \quad  i=1,2.
\end{align}
\end{subequations}
The system matrices are defined as follows:
\begin{align*}
    A{=}\begin{bmatrix}
        F & \mathbf{0}_{4 \times 4} \\
        \mathbf{0}_{4 \times 4} & F
        \end{bmatrix}, \
        F &{=} \begin{bmatrix}
        1      & 5e{\text{-}}2   & 8.8e{\text{-}}3 & 1e{\text{-}}4  \\
        0      & 1      & 3.5e{\text{-}}1   & 8.8e{\text{-}}3  \\
        0      & 0      & 1      & 5e{\text{-}}2    \\
        0      & 0      & 0      & 1
    \end{bmatrix}, \\
    B {=} \begin{bmatrix}
        G & \mathbf{0}_{4 \times 1} \\
        \mathbf{0}_{4 \times 1} & G
        \end{bmatrix}, \
        G&{=} \begin{bmatrix}
        0 & 1e{\text{-}}4 & 1.3e{\text{-}}3 & 5e{\text{-}}2
        \end{bmatrix}^\top.
\end{align*}
Finally, the prediction and control horizon are $T=N=90$. 

In this numerical study, the ball is assumed to be initially stationary and positioned at the center of the plate. Then, we adopt a reference trajectory consisting of five vertices, forming a star shape, which are contained within a circle with $8$-cm radius; we point out that some portions of this trajectory are located outside the constraints{---which prevents us from solving the problem using a traditional tracking MPC}. We consider the following two scenarios regarding the control objective.

\subsection{Scenario 1 -- Pure path-following problem}
Here, only a path-following problem is considered, hence the performance index $\ell^e_{k+j}$ is:
\begin{equation}\label{eq:cost_sc1}
    \mathcal{\ell}^\ellOF\rmFarg{{k}}{\xr},{\ur},{\xr}^{r}) = \lVert {\xr} - {\xr}^{r}  \rVert^2_{E_{\xr}},
\end{equation}
\noindent where ${\xr}^{r}$ is the periodic trajectory reference that the ball has to follow, and $E_{\xr}=\mathrm{diag}(700,0,0,0,700,0,0,0)$, i.e., only the states corresponding to the position of the ball in the plate are weighted. 
Finally, for the stage cost $S$, the identity matrices $Q=10 I_8$ and $R=I_8$ are considered.

\begin{figure}
    \centering
    \includegraphics[width=\columnwidth,trim=0 0.25cm 0 0,clip]{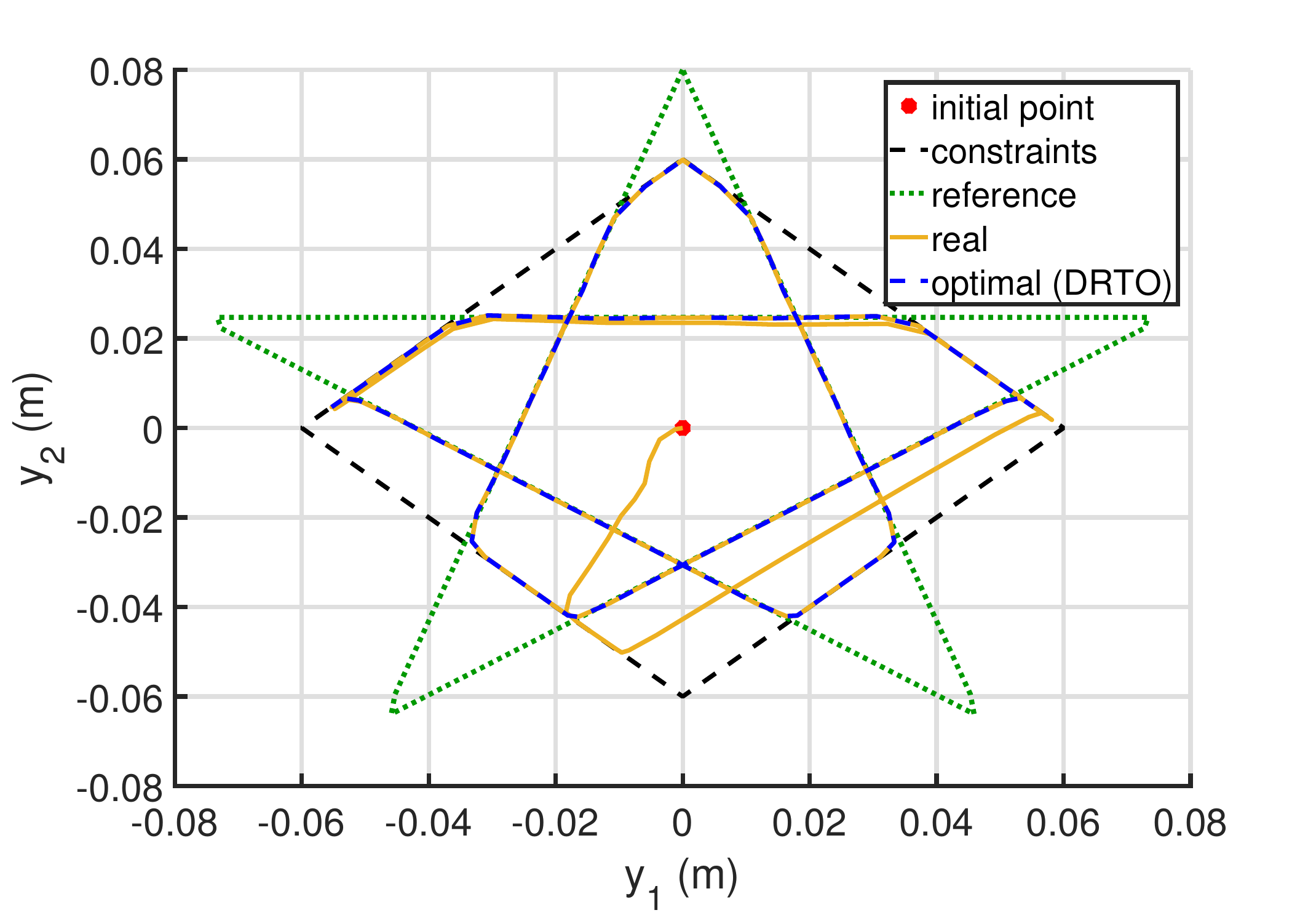}
    \caption{Pure path-following problem. Ball and plate system controlled through Algorithm~\ref{alg:approximatedMPC}. A star-shaped periodic trajectory, which is unreachable in some places, is set as a reference. The ball converges to the optimal periodic trajectory obtained by solving the DRTO problem~\eqref{eq:D}.}
    \label{fig:experiment1_position}
\end{figure}

The simulation results relative to this scenario are shown in Fig.~\ref{fig:experiment1_position}. It can be noticed how the system reaches the target trajectory---without violating the constraints---and converges to the same optimal periodic trajectory obtained by solving the DRTO problem~\eqref{eq:D}.

\begin{figure}
    \centering
    \includegraphics[width=\columnwidth,trim=0 0.25cm 0 0,clip]{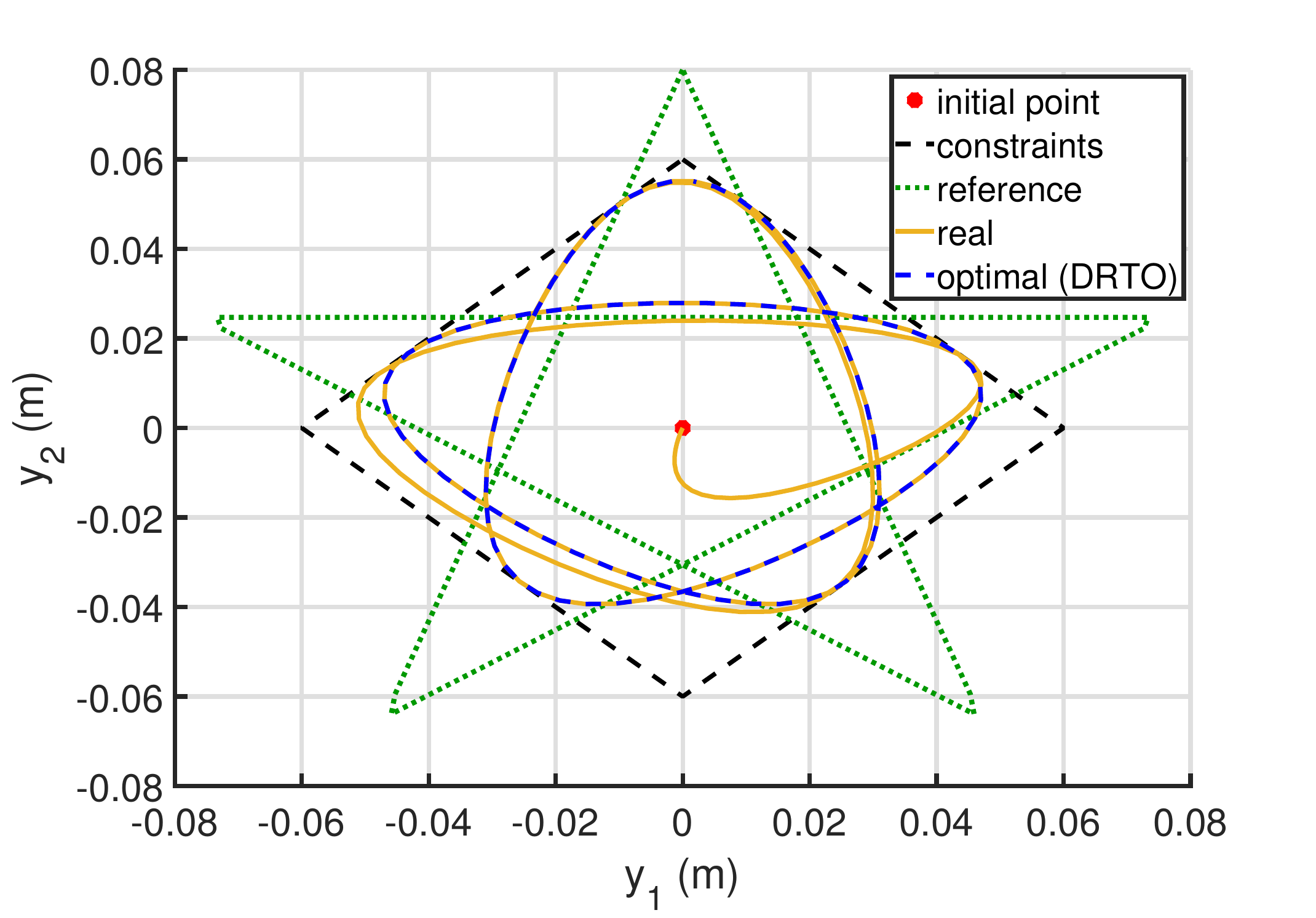}
    \caption{Economic MPC problem. Ball and plate system controlled by Algorithm~\ref{alg:approximatedMPC}. A star-shaped periodic trajectory---partly unfeasible---is set as a reference (same as in Fig.~\ref{fig:experiment1_position}). In this case, the economic objective is to minimize the energy demand of the actuators (hence the acceleration of the plate). The ball converges to the same optimal periodic trajectory obtained by solving the DRTO problem \eqref{eq:D}.}
    \label{fig:experiment2_economic}
\end{figure}
\subsection{Scenario 2 -- Economic optimization problem}

This scenario differs from Scenario~1 in that the economic performance index is also used to measure (and penalize) the energy consumption of the motor.
Assuming that the inertia of the plate remains constant (i.e., the mass of the ball can be neglected), the torque of the plate is proportional to the angular acceleration. Thus, the consumption can be calculated as a function of the acceleration, yielding the following economic performance index:
\begin{equation}\label{eq:cost_sc2}
    \mathcal{\ell}^\ellOF\rmFarg{{k}}{\xr},{\ur},{\xr}^{r}) = \underbrace{\lVert {\xr} {-} {\xr}^{r}  \rVert^2_{E_{\xr}}}_{\text{Reference}} 
    +
    \underbrace{\sum_{i=1}^2 \left(  {a}{\ur}_{i}^2 {-} {b}{\ur}_{i}^4 {+} {c}{\ur}_{i}^6\right )}_{\text{Motor consumption}},
\end{equation}
\noindent where ${\xr}^{r}$ and $E_{\xr}$ are as in Scenario~1, ${a}={c} = 4000$ and ${b} = 6800$. All other parameters remain as in Scenario~1.

Fig.~\ref{fig:experiment2_economic} shows the simulation results relative to this scenario.  It can be observed that here the ball follows a curve with a gentler slope, indicating that implementation of the economic objective results in a limited acceleration input compared to the pure path-following scenario (see Fig.~\ref{fig:experiment1_position}). Analogous conclusions can be drawn about minimizing the distance from the target trajectory while remaining inside the feasible domain, and the convergence to the optimal periodic trajectory of the DRTO problem \eqref{eq:D}.

\subsection{Comparative Analysis}

\begin{figure}
    \centering
    \includegraphics[width=\columnwidth,trim=0 0 1cm 0,clip]{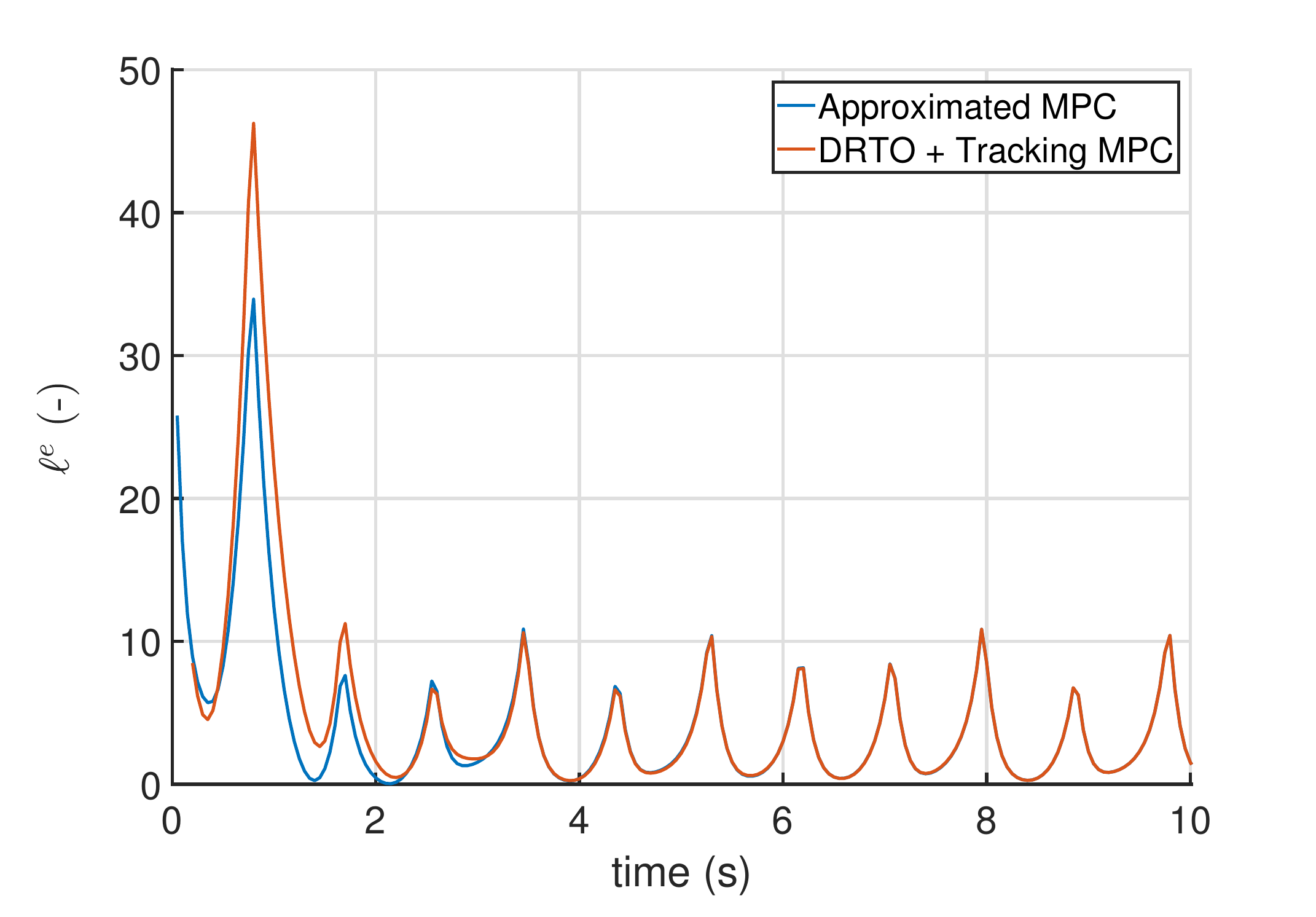}
    \caption{Comparison between the proposed single-layer periodic E-MPC, and a standard MPC for periodic reference tracking (see~\cite{periodictracking_limon}), in following the optimal DRTO periodic trajectory $({\xr}^\optB,\mathbf{\ur}^\optB)$ determined by setting \eqref{eq:Da} as per \eqref{eq:cost_sc2}. Both schemes converge to the same periodic cost evolution (as expected); however, the proposed single-layer E-MPC achieves a more economic solution in the sense of \eqref{eq:cost_sc2} (achieving an average cost of $3.46$ against the $3.71$ for the tracking MPC).}
    \label{fig:experiment2_economic_vs_cost}
\end{figure}

We conclude our numerical study by comparing our single-layer periodic E-MPC scheme with a standard MPC for periodic reference tracking (see~\cite{periodictracking_limon}).
For both controllers, all parameters are set as per the economic case in scenario 2, except for $\ell^e$, which is only used for the E-MPC and set equal to~\eqref{eq:cost_sc2}.
For the tracking MPC, instead, the distance from the given reference trajectory $({\xr}^\optB,\mathbf{\ur}^\optB)$ is penalized via the stage cost matrices $Q$ and $R$, defined as in scenario 1.
Fig.~\ref{fig:experiment2_economic_vs_cost} shows the evolution of the (economic) performance index \eqref{eq:cost_sc2} for both the single-layer E-MPC and the two-layer tracking MPC while the optimal DRTO trajectory is followed.

\section{Conclusions}
\label{sec:conclusions}

In this work, we proposed a single-layer economic model predictive control {design, based on a} QP formulation, for the optimal periodic operation of constrained linear systems, accommodating potential variations in the economic cost function.  The proposed controller  only requires knowledge of the gradient of the economic cost function, which is  updated online as described in our Algorithm~\ref{alg:approximatedMPC}.  This allows for easy implementation on industrial platforms through the use of quadratic solvers. We prove feasibility, stability and convergence to optimum of the proposed approach. In addition, its performance {is  validated on the} control of a ball-and-plate system. This shows how the controller is capable of optimally following a specified periodic trajectory while simultaneously minimizing a {given} economic cost.


\end{document}